\documentclass{article}

\begin{document}

{\Large\bf
\begin{center}{Cryptanalysis of  a Practical \hbox{Quantum Key Distribution} With Polarization-Entangled Photons}
\end{center}}

\centerline{Thomas Beth, J\"orn M\"uller-Quade, and Rainer Steinwandt}

\begin{center}{IAKS/E.I.S.S., Fakult\"at f\"ur Informatik,
Am Fasanengarten 5,\\Universit\"at Karlsruhe (TH),
76131 Karlsruhe, Germany}
\end{center}

\begin{abstract}{\noindent
Recently, a quantum key exchange protocol has
been described \cite{PFLM04}, which served as basis for securing an actual bank
transaction by means of quantum cryptography \cite{ZVS04}.
Here we show, that the authentication scheme applied is insecure in the sense that an attacker can
provoke a situation where initiator and responder of a key exchange end up with
different keys. Moreover, it may happen that an attacker can
decrypt a part of the plaintext protected with the derived encryption key.}
\end{abstract}


\section{Introduction}
\noindent In April 2004, in Vienna an actual bank transfer was protected by means of a one-time-pad-based encryption \cite{ZVS04} where the
one-time-pad has been derived by means of a quantum key
exchange using a novel authentication scheme.
However, as pointed out, e.\,g., by Raub et
al. \cite{RSM04}, using the ``textbook version'' of a one-time-pad for encrypting a bank
transfer is not a suitable choice, if the plaintext involves no further integrity
protection: assume, for instance, the amount of money to be transferred is represented as an ASCII
string which is XORed with the one-time-pad. Then, by just flipping certain bits in the
ciphertext, an attacker may change the amount of money to be
transferred. Similarly, the attacker may be able to change the name of the recipient of the
money. Thus, the one-time-pad encryption should be combined with (unconditionally secure)
means ensuring the authenticity of the plaintext \cite{RSM04}. However, even a scheme
modified in this sense would not provide a secure bank transfer:

In this contribution we describe an attack on the quantum key exchange scheme itself
that has been used in the Vienna experiment. Due to a flaw in the
classical authentication part, an attacker may gain access
to a part of the plaintext later encrypted under an established key. Also she may provoke a
situation where the participants of the key exchange end up with different keys without
noticing this. Of course, a trivial denial-of-service-attack (``cutting all wires'') may
also prevent the users from establishing a shared key; but the attack presented here is more severe in the
sense that both protocol participants obtain a key which they might bring to use, as---differing from the ``cut the wires approach''---the failure of the key exchange remains
undetected.

\section{The quantum key exchange scheme used in Vienna}
\noindent The published version of the quantum key exchange protocol does not describe all details
of the Vienna experiment. However, for describing our attack this is not really necessary, and
it is sufficient to look at the (classical) privacy amplification and authentication part. 
Owing to the attacks described below, the published version of these parts of the protocol \cite{CamerinoPaper} deviates from the version used
in the Vienna experiment; from the latter at the time of writing only a poster presentation \cite{PML04} was available to us, and we are indebted to Momtchil Peev for kindly providing us with further details \cite{Pee04}.
In summary,
for establishing a common key between Alice and Bob, the following steps are
performed:
\begin{itemize}
  \item A raw key between Alice and Bob is established by means of polarization-entangled
        photons.
  \item In a sifting step, parts of the raw key are discarded based on a public discussion
        between Alice and Bob.
  \item Next, the quantum bit error rate is estimated based upon which the protocol is
        either aborted or continued with an error correction step.
  \item Hereafter, privacy amplification is performed, based on a matrix sent from Alice
        to Bob. The result of the privacy amplification is the final key, if the subsequent
        authentication step succeeds.
  \item Finally, a \emph{protocol-log extract} is formed from the messages sent throughout
        the protocol so far; the authenticity of this log is checked by means of a message
        authentication procedure. The final key from the privacy amplification phase is
        accepted if and only if this authentication check succeeds.
\end{itemize}
As already indicated above, for our attack only the last two steps are of importance, as
only one variant of our  attack interferes with the quantum part of the protocol.
\paragraph{Privacy amplification} This part of quantum key exchange protocols has been
introduced by Bennett et al. \cite{BBR88} and is based on a binary 
rectangular matrix $P_A$ with random entries. Multiplying $P_A$ with the raw key yields the
shorter final key about which the adversary has only negligible information. Thus, each
row of $P_A$ determines one bit of the final key.
\paragraph{Protocol-log extract} The protocol-log extract is comprised of five parts (and
has to be identical for Alice and Bob) \cite{Pee04}:
\begin{itemize}
   \item the basis for each sifted bit;
   \item the positions of the bits disclosed in the process of error estimation;
   \item the estimated error rate;
   \item the positions of the bits corrected by the specific error correction routine;
   \item the last 128~bits of the jointly generated key (these are subsequently discarded).
\end{itemize}
Note that the only part of the protocol-log extract influenced by the privacy amplification matrix are the
last 128~bits of the jointly generated key; the privacy amplification matrix itself is not explicitly included, and
the attack described in the next section exploits this.

\section{Cryptanalysis of the scheme}\label{sec:attack}
\noindent By construction of the protocol, only 128 rows of the privacy amplification matrix $P_A$
affect the protocol-log transcript. The remaining rows of $P_A$ remain unauthenticated
and hence can be modified by the attacker at will. E.\,g., she can
\begin{itemize}
   \item replace all (but the last 128) rows of $P_A$ by random vectors. Consequently, the receiver
   Bob of $P_A$ will end up with a key different from Alice's, but neither Alice nor Bob
   is aware of this fact. In particular, later bringing this key to use may result in the
   failure of an application---even when the attacker is not interfering with that application;
   \item flip an individual entry in the $i$-th row of $P_A$. Then with a success
   probability of ca. 0.5 (namely, if the corresponding bit in the raw key is set) she can
   flip the $i$-th bit of the key derived by Bob.
\end{itemize}
Now suppose that the attacker succeeds in measuring a small number of qubits---logarithmic in the total number of
qubits sent---and assume further that few qubits of the sifted key after error correction are known to the
attacker (this happens with polynomial probability).
Then the attacker may proceed as follows: She replaces one row of the privacy amplification
matrix with a binary vector containing ones only at positions corresponding to bits of the
sifted and error-corrected key she knows. In this way she learns a bit of the final key
derived by Bob. Consequently, if the key later is used to encrypt a message from Bob to
Alice by a bit-wise XOR, then the attacker immediately learns the respective plaintext
bit.
In fact, in the proposed form of the key exchange protocol, the attacker may use a trivial method for
learning the complete key derived by Bob: even replacing---up to the last 128---all rows of the privacy
amplification matrix by zero vectors remains undetected and results
in the all-zero key for Bob.

\section{Including the privacy amplification matrix in\-to the protocol-log extract}
\noindent From the above discussion it may be tempting to conclude that including the complete privacy
amplification matrix into the protocol-log extract is sufficient for securing the protocol. However, to show that this approach does not offer acceptable cryptographic security let us consider a variant of the above protocol in which the complete matrix $P_A$ is
included in the protocol-log extract and authenticated. (We stress
that this variant of the protocol has not been proposed or used for
the Vienna experiment \cite{PFLM04}).

Let us recall the authentication procedure applied
in the Vienna experiment: For authenticating the protocol-log extract $M$,
first it is compressed by a publically known cryptographic hash
function $H_0$ like SHA-256, and for all subsequent
computations $M$ is identified with its hash value under $H_0$. However, in the presence
of an unlimited adversary such an identification does not rule out the following attack:
\begin{itemize}
   \item The attacker impersonates Bob and follows the quantum key exchange up to the
         point where
         Alice sends the authenticated hash $H_0(M_A)$ of her protocol-log extract $M_A$. Here
         the attacker aborts the protocol with Alice.
   \item Now, the attacker impersonates Alice and initiates a quantum key exchange with
         Bob. The attacker follows the protocol up to the point where the privacy
         amplification matrix $P_A$ is to be chosen.
   \item Instead of choosing a random $P_A$, she
         makes an exhaustive search over all possible matrices of the
         appropriate size to find a matrix $P_A'$ which, when included in the protocol-log
         extract, yields the same hash value $H_0(M_A)$ as obtained from Alice. Such a
         $P_A'$ exists with overwhelming probability, if we model $H_0(\cdot)$ as a random oracle.
         As there are significantly more degrees of freemdom in the privacy amplification matrix
         than in the typical output of a a cryptographic hash function (like, e.\,g.,
         SHA-256), the existence of such a $P_A'$ is plausible.

         For actually performing this exhaustive search, the attacker exploits that up to the last
         128 bits of the final key (which only depend on the data collected
         so far and the privacy amplification matrix), the protocol-log extract is
         completely known.
   \item The privacy amplification matrix $P_A'$ along with the authenticated hash $H_0(M_A)$ obtained from
         Alice are sent to Bob, who will accept this as a valid authentication.
   \item The subsequent authentication information from Bob is ignored, and the attacker
         can impersonate Alice in the subsequent use of the final key.
\end{itemize}\medskip

\noindent To avoid the above attacks and ensure that the privacy
amplification matrix is identical for Alice and Bob, Peev et
al. \cite{CamerinoPaper} make use of a scheme of Gilbert and Hamrick \cite{GiHa04} where the
privacy amplification matrix is not sent over the public channel but derived from
previously authenticated data.

\section{Conclusions}
\noindent The above discussion shows that in the original form the quantum key exchange scheme used
in the Vienna protocol \cite{PML04,PFLM04} does not offer acceptable
cryptographic security. Similarly as the ``malleability problem'' pointed out by Raub et
al. \cite{RSM04}, our attacks focus on the classical parts of the protocol and provide
evidence of the importance of classical cryptographic aspects in quantum cryptography.

\section*{Acknowledgements}
\noindent
We are indebted to Momtchil Peev for his detailed explanations of the authentication step of the quantum key exchange protocol.
Further on, this work was partially supported by the DFG project
ANTI-BQP and the projects \hbox{{\sc ProSecCo} (IST-2001-39227)}
and SECOQC of the European Commission.


\begin{thebibliography}{000}
\bibitem{BBR88}
C.H. Bennett, G. Brassard, and J.-M. Robert.
\newblock{Privacy amplification by public discussion}.
\newblock {SIAM Journal on Computing, Vol. 17, pp. 210--229}, 1988.

\bibitem{GiHa04}
G. Gilbert and M. Hamrick.
\newblock{Practical Quantum Cryptography: A Comprehensive Analysis (Part One)}.
\newblock {LANL-preprint quant-ph/0009027}, 2004.

\bibitem{Pee04}
M. Peev.
\newblock{Personal email communication}, 2004.

\bibitem{PML04}
M. Peev, O. Maurhardt, T. Lor\"unser, M. Suda, M. N\"olle, A. Poppe, R. Ursin, A. Fedrizzi, H. B\"ohm,
T. Jennewein, and A. Zeilinger
\newblock{A Novel Protocol-Authentication Algorithm in Quantum Cryptography}.
\newblock{Poster presentation at International Meeting on Quantum Information Science Foundations of Quantum Information,
University of Camerino, Camerino, Italy}, 2004. Preliminary version of the paper by Peev et al. \cite{CamerinoPaper}.

\bibitem{CamerinoPaper}
M. Peev, M. N\"olle, O. Maurhardt, T. Lor\"unser, M. Suda, A. Poppe, R. Ursin, A. Fedrizzi, and A. Zeilinger
\newblock{A Novel Protocol-Authentication Algorithm Ruling Out a Man-in-the Middle Attack in Quantum Cryptography}.
\newblock{Submitted for publication}, 2004.

\bibitem{PFLM04}
A. Poppe, A. Fedrizzi, T. Lor\"unser, O. Maurhardt, R. Ursin, H.R. B\"ohm, M. Peev,
M. Suda, C. Kurtsiefer, H. Weinfurter, T. Jennewein, and A. Zeilinger.
\newblock{Practical Quantum Key Distribution with Polarization-Entangled Photons}.
\newblock {LANL-preprint quant-ph/0404115}, 2004.

\bibitem {RSM04}
D. Raub, R. Steinwandt, and J. M\"uller-Quade.
\newblock{On the Security and Composability of the One Time Pad}.
\newblock{Cryptology ePrint Archive: Report 2004/113}, 2004.

\bibitem{ZVS04}
A collaboration of: group of Prof. Anton Zeilinger, Vienna University; ARC Seibersdorf research GmbH; City of Vienna; Wien Kanal Abwassertechnologien GmbH and Bank Austria -- Creditanstalt.
\newblock{Quantum Cryptography ``live'', {\tt http://www.quantenkryptographie.at/rathaus\_press.html}}, 2004.

\end{thebibliography}
\end{document}